\documentclass[11pt]{article}

\usepackage{amsmath}
\usepackage{epsfig}
\usepackage{graphics}

\newcommand{\N}{N\raise.7ex\hbox{\underline{$\circ $}}$\;$}

\begin{document}

\title{
E.M. Ovsiyuk\footnote{e.ovsiyuk@mail.ru}, V.M. Red'kov\footnote{redkov@dragon.bas-net.by}\\
On simulating a medium  with special reflecting properties\\  by Lobachevsky geometry
\\ (One exactly solvable  electromagnetic  problem) }

\maketitle

\begin{quotation}
Lobachewsky geometry simulates a  medium with special consti\-tutive relations,
$
D^{i} = \epsilon_{0} \epsilon^{ik} E^{k},  B^{i} =
\mu_{0} \mu^{ik} H^{k}$,
where two matrices coincide:
$
\epsilon^{ik}(x) = \mu^{ik}(x)
$. The situation is specified in quasi-cartesian coordinates $(x,y,z)$.
Exact solutions of the Maxwell equations in  complex  3-vector ${\bf E}+i {\bf B}$ form, extended to curved
space models within the tetrad formalism, have been found in Lobachevsky space.
The problem reduces to a second order diffe\-rential equation which can be associated with an 1-dimensional
Schr\"{o}dinger problem for a particle in external  potential field $U(z) =  U_{0}e^{2z}$.
In quantum mechanics, curved geometry acts as an effective potential barrier with reflection coefficient $R=1$;
in electrodynamic  context  results similar to quan\-tum-mechanical  ones   arise:
the Lobachevsky  geometry   simu\-lates  a  medium  that effectively  acts as
an ideal  mirror.
Penetration of the electromagnetic field into the effective medium, depends on the parameters of an
electromagnetic wave,
frequency $\omega, \; k_{1}^{2} + k_{2}^{2}$, and the curvature radius $\rho$.

\end{quotation}

\section{Introduction}

An aim of the present paper is to obtain exact solutions of the Maxwell
equations in 3-dimensional Lobachevsky space $H_{3}$. A coordinate system used
is one from the list given by Olevsky \cite{Olevsky}, which generalizes
  Cartesian coordinate in flat Euclidean space.

  To treat Maxwell equations we make use of complex representation of these according to the known approach
  by Riemann--Silberstein--Oppenheimer--Majorana  \cite{1901-Weber, 1907-Silberstein(1), 1931-Oppenheimer,
1931-Majorana}  (see also in  [6 -- 30]),  extended to curved space-time models in the  frames of tetrad
  formalism of Tetrode--Weyl-Fock--Ivanenko  \cite{1928-Tetrode, 1929-Weyl(1), 1929-Fock-Ivanenko};
   see also in  \cite{Book}).
On the base of this technique, new exact solutions of the type of extended plane wave in Lobachevsky space
have been constructed explicitly. These may be interesting in the cosmological sense; besides, they  may be interesting in
the context
 of geometric simulating electromagnetic field in a special medium
 \cite{1973-Landau}, \cite{Book}.

\section{ Cartezian coordinates in Lobachevsky space}

In Olevsky paper \cite{Olevsky}, under the number 2
the following coordinate system in Lobachevsky space $H_{3}$ is specified
\begin{eqnarray}
x^{a} = (t , x, y, z) \; , \qquad dS^{2}= dt^{2} - e^{-2z} (
dx^{2} + dy^{2} ) - dz^{2} \; ,
\label{2.1}
\end{eqnarray}

\noindent   the element of volume is given by
\begin{eqnarray}
 dV =\sqrt{-g} \; dx dy dz = e^{-2z} dx dy dz \;, \qquad  x,y,z \in ( - \infty , + \infty )\; ;
\nonumber
\end{eqnarray}

\noindent     the magnitude and sign of the  $z$   are substantial, in particular, when dealing with
localization, for example, the energy of the field
\begin{eqnarray}
dW = {1 \over 2} ({\bf E}^{2} + {\bf B}^{2} )d V =   {1 \over 2} ({\bf E}^{2} + {\bf B}^{2} ) \; e^{-2z}  \; dx dy dz \; .
\label{2.2}
\end{eqnarray}

It is helpful  to have at hand  some detail of the parametrization of the model $H_{3}$ by
 $x,y,z)$. It is known that this model can be identified with a branch of hyperboloid in 4-dimension flat space
\begin{eqnarray}
u_{0}^{2} - u_{1}^{2} - u_{2}^{2} - u_{3}^{2} = \rho^{2} \; , \qquad
u_{0} = + \sqrt{\rho^{2}  + {\bf u}^{2} } \;  .
\nonumber
\end{eqnarray}

\noindent Coordinate in use,   $x,y,z$, are referred to   $u_{a}$ by relations
\begin{eqnarray}
u_{1}=x e^{-z}\; , \;\; u_{2}=y e^{-z}\; , \;\;
\nonumber
\\
u_{3}={1\over 2}[(e^{z}-e^{-z})+(x^{2}+y^{2})e^{-z}]\; ,
\nonumber
\\
u_{0}={1\over 2}[(e^{z}+e^{-z})+(x^{2}+y^{2})e^{-z}]\; .
\label{2.3}
\end{eqnarray}

\noindent It is convenient to employ 3-dimensional Poincar\'{e} realization for
Lobachevsky  space as  inside part of 3-sphere
\begin{eqnarray}
q_{i} = {u_{i} \over  u_{0}}
 = {u_{i} \over  \sqrt{\rho^{2} + u_{1}^{2} + u_{2}^{2} + u_{3}^{2} }}, \qquad q_{i}q_{i} < +1 \; .
\label{2.4}
\end{eqnarray}

Quasi-Cartesian coordinates $(x,y,z)$  are referred to  $q_{i}$ as follows
\begin{eqnarray}
q_{1} = {2 x \over  x^{2} + y^{2}  + e^{2z} + 1}\;,
\nonumber
\\
q_{2} = {2 y \over  x^{2} + y^{2}  + e^{2z} + 1}\; ,
\nonumber
\\
q_{3} = {x^{2} + y^{2} + e^{2z} -1 \over z^{2} + y^{2}  + e^{2z} +
1} \; ;
\label{2.5}
\end{eqnarray}

Inverses to (\ref{2.5}) relations are
\begin{eqnarray}
x =  {q_{1} \over 1 - q_{3}} \; , \qquad y =    {q_{2} \over 1 -
q_{3}} \; , \qquad e^{z} =  { \sqrt{1 - q^{2}} \over 1 - q_{3}}
\; .
\label{2.6}
\end{eqnarray}

In particular, note that on the  axis $q_{1}=0, q_{2}=0, q \in (-1 , +1)$ relations (\ref{2.6})
assume the form
\begin{eqnarray}
x =  0 \; , \qquad y = 0 \; , \qquad e^{z} =   \sqrt{ {1 + q_{3}
\over 1 - q_{3}}  }    \; .
\nonumber
\end{eqnarray}

\noindent
that is
\begin{eqnarray}
q_{3} \longrightarrow  +1 \; , \qquad e^{z} \longrightarrow + \infty \; , \qquad z \longrightarrow  + \infty \; ;
\nonumber
\\
q _{3} \longrightarrow  -1 \; , \qquad e^{z} \longrightarrow  + 0 \; ,
\qquad  z \longrightarrow - \infty \; .
\label{2.7}
\end{eqnarray}

Solutions of the Maxwell equation, constructed bellow, can be of interest in the context of
description of electromagnetic waves in special media, because the Lobachevsky geometry
simulates effectively a definite special medium \cite{2009-Red'kov-Tokarevskaya-Ovsiyuk-Spix},
 inhomogeneous along  the axis $z$.
Effective electric permittivity  tensor  $\epsilon^{ik}(x)$ is given by
\begin{eqnarray}
\epsilon^{ik}(x) = -\sqrt{-g} \; g^{00} (x) g^{ik}(x) =\left | \begin{array}{ccc}
  1  &0 & 0\\
0 &  1 & 0\\
0 & 0 & e^{-2z}
\end{array}  \right |,
\label{2.8}
\end{eqnarray}

\noindent whereas the corresponding effective magnetic  permittivity tensor is
 \begin{eqnarray}
(\mu^{-1}) ^{ik}(x) =  \sqrt{-g} \left | \begin{array}{ccc}
g^{22}g^{33} & 0 &  0\\
0 & g^{33}g^{11} & 0\\
0  & 0 &g^{11}g^{22}
\end{array} \right |= \left | \begin{array}{ccc}
1 & 0 &  0\\
0 & 1 & 0\\
0  & 0 &e^{2z}
\end{array} \right |.
\label{2.9}
\end{eqnarray}

\noindent In explicit form, effective constitutive relations  (the system {\em SI} is used) are
\begin{eqnarray}
D^{i} = \epsilon_{0} \epsilon^{ik} E_{k} \; ,   \qquad B_{i} =
\mu_{0} \mu^{ik} H^{k} \; ,
\label{2.9}
\end{eqnarray}

\noindent note that two matrices coincide:
$
\epsilon^{ik}(x) = \mu^{ik}(x)
$.

\section{Tetrads and Maxwell equations in complex form }

In the coordinate (\ref{2.1}), let us introduce  a tetrad
\begin{eqnarray}
e_{(a)}^{\beta} = \left | \begin{array}{cccc}
1 & 0 & 0 & 0 \\
0 & e^{z} & 0 & 0 \\
0 & 0 & e^{z} & 0 \\
0 & 0 & 0 & 1
\end{array} \right | \; , \qquad
 e_{(a) \beta} = \left |
\begin{array}{llll}
1 & 0 & 0 & 0 \\
0 & -e^{-z} & 0 & 0 \\
0 & 0 & - e^{-z} & 0 \\
0 & 0 & 0 & - 1
\end{array} \right | \; .
\label{3.1}
\end{eqnarray}

One should find Christoffel symbols;
some of them evidently vanish:
$
\Gamma^{0}_{\beta \sigma} = 0 \; , \; \Gamma^{i}_{00} = 0 \; ,
\; \Gamma^{i}_{0j} = 0$, remaining ones are determined by relations
\begin{eqnarray}
\Gamma^{x}_{\;\; jk} = \left | \begin{array}{ccc}
0 & 0 & -1 \\
0 & 0 & 0 \\
-1 & 0 & 0
\end{array} \right |  ,
 \Gamma^{y}_{\;\; jk} = \left | \begin{array}{ccc}
0 & 0 & 0 \\
0 & 0 & -1 \\
0 & -1 & 0
\end{array} \right |  ,
\Gamma^{z}_{\;\; jk} = \left | \begin{array}{ccc}
e^{-2z} & 0 & 0 \\
0 & e^{-2z} & 0 \\
0 & 0 & 0
\end{array} \right |  .
\nonumber
\end{eqnarray}

\noindent Ricci rotation coefficients are (only not vanishing ones are written down)
\begin{eqnarray}
 \gamma_{31 1}  = -1\; , \qquad \gamma_{23 2}
= 1 \; .
\nonumber
\end{eqnarray}

Using the notation  \cite{Book}
\begin{eqnarray}
e_{(0)}^{\rho} \partial_{\rho} = \partial_{(0)} = \partial_{t} \;
, \qquad e_{(1)}^{\rho} \partial_{\rho} = \partial_{(1)} =
e^{z}\partial_{x}\; , \nonumber
\\
e_{(2)}^{\rho} \partial_{\rho} = \partial_{(2)} = e^{z}
\partial_{y} \; , \qquad e_{(3)}^{\rho} \partial_{\rho} =
\partial_{(3)} = \partial_{z} \; ,
\nonumber
\\
 {\bf v}_{0} =( \gamma_{01 0}, \gamma_{02 0} , \gamma_{03 0} ) \equiv 0 \; , \qquad
 {\bf v}_{1} = ( \gamma_{01 1}, \gamma_{021 } , \gamma_{03 1} ) \equiv 0 \; ,
\nonumber
\\
 {\bf v}_{2} = ( \gamma_{012 0}, \gamma_{02 2} , \gamma_{03 2} ) \equiv 0 \; ,
\qquad
 {\bf v}_{3} = ( \gamma_{013}, \gamma_{02\; 3} , \gamma_{03 3} ) \equiv 0 \; ,
\nonumber
\\
 {\bf p}_{0} = ( \gamma_{23 0}, \gamma_{31 0} , \gamma_{12 0} ) = 0 \; , \qquad
 {\bf p}_{1} = ( \gamma_{23 1}, \gamma_{31 1} , \gamma_{12 1} ) = (0 , -1, 0) \; ,
 \nonumber
\\
 {\bf p}_{2} = ( \gamma_{23 2}, \gamma_{31 2} , \gamma_{12 2} ) = (1 , 0, 0) \; ,
\qquad
 {\bf p}_{3} = ( \gamma_{23 3}, \gamma_{31 3} , \gamma_{12 3} ) = 0
\; ;
\nonumber
\end{eqnarray}

\noindent
the Maxwell equations in the  complex matrix form   \cite{Book} read
\begin{eqnarray}
\left [ \; \alpha^{k} \; \partial_{(k)} + {\bf s} {\bf v}_{0} +
 \alpha^{k} \; {\bf s} {\bf p}_{k}
-
 i\;
 ( \; \partial_{(0)} +
  {\bf s} {\bf p}_{0} - \alpha^{k} {\bf s} {\bf v}_{k} ) \right ] \left | \begin{array}{c}
0 \\ {\bf E} + i {\bf B}
\end{array} \right |
 = 0\; ;
\label{3.2}
\end{eqnarray}

\noindent in the used retrad it assumes the form
\begin{eqnarray}
\left ( - i \partial_{t} + \alpha^{1}  e^{z} \partial_{x} +
\alpha^{2}  e^{z}  \partial_{y} +
 \alpha^{3} \partial_{z}
- \alpha^{1}  s_{2}  +\alpha^{2}  s_{1} \right )
 \left | \begin{array}{c}
0 \\ {\bf E} + i {\bf B}
\end{array} \right | = 0 \; .
\label{3.3}
\end{eqnarray}

\noindent Matrices involved in (\ref{3.3}) are
\begin{eqnarray}
\alpha^{1} = \left | \begin{array}{rrrr}
0 & 1 & 0 & 0 \\
-1 & 0 & 0 & 0 \\
0 & 0 & 0 & -1 \\
0 & 0 & 1 & 0
\end{array} \right |  ,\qquad
\alpha^{2} = \left | \begin{array}{rrrr}
0 & 0 & 1 & 0 \\
0 & 0 & 0 & 1 \\
-1 & 0 & 0 & 0 \\
0 & -1 & 0 & 0
\end{array} \right |  ,\nonumber
\\
 \alpha^{3} = \left | \begin{array}{rrrr}
0 & 0 & 0 & 1 \\
0 & 0 & -1 & 0 \\
0 & 1 & 0 & 0 \\
-1 & 0 & 0 & 0
\end{array} \right | ,
s^{1}= \left | \begin{array}{rrrr}
0 & 0 & 0 & 0 \\
0 & 0 & 0 & 0 \\
0 & 0 & 0 & -1 \\
0 & 0 & 1 & 0 \\
\end{array} \right | ,
s^{2} = \left | \begin{array}{rrrr}
0 & 0 & 0 & 0 \\
0 & 0 & 0 & 1 \\
0 & 0 & 0 & 0 \\
0 & -1 & 0 & 0 \\
\end{array} \right | .
 \nonumber
\end{eqnarray}

\section{Separation of the variables }

Let us use the substitution
\begin{eqnarray}
\left | \begin{array}{c} 0 \\ {\bf E} + i {\bf B}
\end{array} \right | = e^{-i \omega t } \; e^{i k_{1} x} \; e^{ik_{2} y} \left | \begin{array}{c}
0 \\ {\bf f}(z)
\end{array} \right | \;  .
\label{4.1}
\end{eqnarray}

\noindent correspondingly eq.  (\ref{4.1}) gives
\begin{eqnarray}
\left  ( - \omega + \alpha^{1}  e^{z} i k_{1} + \alpha^{2}
e^{z} i k_{2} +
 \alpha^{3} {d \over dz}
- \alpha^{1}  s_{2}  +\alpha^{2}  s_{1} \right  ) \left
|
\begin{array}{c} 0 \\ f_{1}(z) \\ f_{2}(z) \\ f_{3}(z)
\end{array} \right |
 = 0 \; .
\label{4.2}
\end{eqnarray}

After simple calculation,we derive a first order system for  $f_{i}$:
\begin{eqnarray}
i k_{1} \; e^{z}  f_{1} +  i k_{2} \; e^{z} \;
f_{2} + ({d \over d z}-2 ) f_{3} = 0 \; ,
\nonumber
\\
-\omega f_{1} - ( {d \over dz}-1 ) f_{2} +  i k_{2} \; e^{z} \; f_{3} = 0
\; , \nonumber \\
-\omega f_{2} + ({d \over dz}-1)f_{1} -  i k_{1}\; e^{z} \; f_{3} = 0
\; , \nonumber
\\
-\omega f_{3} - e^{z} \;i k_{2} f_{1} +  i k_{1} \; e^{z} \; f_{2} = 0
\; .
\label{4.4}
\end{eqnarray}

\noindent
Allowing  three last equations in the first one, we get an identity
$
 0  = 0 $. So, there exist only three independent equations
  (below the notation $k_{1}= a, k_{2}=b$ is used):
\begin{eqnarray}
\omega f_{3} = -  i b \; e^{z} \; f_{1} +  i a \; e^{z} \; f_{2} \; ,
\nonumber
\\[2mm]
\omega f_{1} =  - ({d \over d z}-1)f_{2} + i b  \;  e^{z} \; f_{3} \; ,
\nonumber
\\
\omega f_{2} = +({d \over d z}-1)f_{1} -i a  \;  e^{z} \; f_{3} \;
,
\label{4.5}
\end{eqnarray}

With substitutions
$
f_{1} = e^{z} F_{1}(z)\; , \;
 f_{2} = e^{z} F_{2}(z)$ ,
eqs.  (\ref{4.5})  give
\begin{eqnarray}
\omega f_{3} = -  i b \; e^{2z} \; F_{1} +  i a \; e^{2z} \; F_{2} \; ,
\nonumber
\\
\omega F_{1} =  - {d \over d z} F_{2} + i b  \;   f_{3} \; ,
\nonumber
\\
\omega F_{2} = + {d \over d z} F_{1} -i a  \;  f_{3} \; .
\label{4.6}
\end{eqnarray}

There exist a particular case readily treatable, when
$a=0, \; b = 0  , \; f_{3} = 0 $:
\begin{eqnarray}
\omega F_{1} =  - {d \over d z} F_{2} \; , \qquad
\omega F_{2} = + {d \over d z} F_{1}  \;  \Longrightarrow
\nonumber
\\
 F_{1} (z) = e^{ \pm i\omega z} \; , \qquad  F_{2} =  \pm i \;  e^{ \pm i\omega z}  \; ,
\label{4.7}
\end{eqnarray}
which gives
\begin{eqnarray}
\Phi^{\pm}  = \left | \begin{array}{c}
0 \\
{\bf E} + i {\bf B}
\end{array} \right | =e^{-i\omega t} e^{z}
\left | \begin{array}{c}
0\\
e^{\pm i \omega z}
\\
\pm i\; e^{\pm i \omega z}
\\
0
\end{array} \right |
\end{eqnarray}
or (let it be $\varphi^{(\pm)}  = \omega t  \mp \omega z $)
\begin{eqnarray}
E_{1} ^{(\pm)}  + i B_{1}^{(\pm)} = \cos ( \omega t  \mp \omega z) - i  \sin (\omega t  \mp \omega z)\; ,
\nonumber
\\
E_{2}^{(\pm)} + i B_{2}^{(\pm)} = \pm  \sin (\omega t  \mp \omega z) \pm i\; \cos (\omega t  \mp \omega z) \;.
\label{4.8}
\end{eqnarray}

\noindent It is easily checked the known presupposed property
$
{\bf E}^{(\pm)} \times {\bf B}^{(\pm)} =  \pm {\bf e}_{z}$.

Let us turn back to the generale case   (\ref{4.6}), from the first equation it follows
\begin{eqnarray}
 f_{3} = {- i b   \over \omega} \;  e^{2z}  F_{1} + {i a  \over \omega} \; e^{2z}
 F_{2}
\; ,
\label{4.10}
\end{eqnarray}

\noindent
and further we get a system for  $F_{1}$ and $F_{2}$
\begin{eqnarray}
({d \over d z}  + {a b \;  e^{2z}\over \omega}) \; F_{2} =
{b^{2}e^{2z}-\omega^{2} \over \omega} \; F_{1} \; , \nonumber
\\
({d \over d z}  - {a b \; e^{2z}\over \omega}) \; F_{1} =
{\omega^{2}-a^{2}e^{2z} \over \omega} \; F_{2} \; .
\label{4.11}
\end{eqnarray}

With the help of a new variable
$
e^{z} =  \sqrt{\omega} \; Z $,  two last are written as
\begin{eqnarray}
Z \; ({d \over d  Z}  +  a b \;  Z)\; F_{2}= +
 (b^{2}  Z^{2}- \omega ) \; F_{1} \; ,
\nonumber
\\
Z \; ({d \over d Z}  - a b \;  Z ) \; F_{1}=
- ( a^{2} Z^{2}  - \omega) \; F_{2} \; .
\label{4.12}
\end{eqnarray}

This system can be solved straightforwardly in terms if Heun confluent functions.
Indeed, from (\ref{4.12}) it follows a second order differential equation
for  $F_{1}$
\begin{eqnarray}
{d^{2}F_{1}\over dZ^{2}}-{a^{2}Z^{2}+\omega\over Z\,(a^{2}Z^{2}-\omega)}\,
{dF_{1}\over dZ}+\left[{\omega^{2}\over Z^{2}}+{2\,ab \,\omega\over a^{2}Z^{2}-\omega}-
(a^{2}+b^{2})\,\omega\right]F_{1}=0\,,
\label{4.13}
\end{eqnarray}

\noindent here we note additional  singular point ar $
Z=\pm \sqrt{\omega} /  a$.
With the new variable, we get
\begin{eqnarray}
y={a^{2}Z^{2}\over \omega}\,, \qquad
{d^{2}F_{1}\over dy^{2}}+\left[{1\over y}-{1\over y-1}\right]{dF_{1}\over dy}
\nonumber\\
+\left[{\omega^{2}\over 4\,y^{2}}-{2\,ab\,\omega+(a^{2}+b^{2})\,\omega^{2}\over 4\,a^{2}y}+{b\omega\over 2a\,(y-1)}\right]F_{1}=0\,.
\label{4.14}
\end{eqnarray}

\noindent from whence or with the  substitution  $F_{1}(y)=y^{c}\,g_{1}(y)$ we arrive at
\begin{eqnarray}
{d^{2}g_{1}\over dy^{2}}+\left[{2c+1\over y}-{1\over y-1}\right]{dg_{1}\over dy}
+\left[{\omega^{2}/4+c^{2}\over y^{2}}\,\right.
\nonumber\\
\left.+\,{2c-\omega^{2}/2-b\,\omega/a-b^{2}\omega^{2}/(2a^{2})\over 2\,y}+{-2c+b\omega/a\over 2\,(y-1)}\right]g_{1}=0\,.
\label{4.15}
\end{eqnarray}

\noindent When  $c=\pm i\omega/2$, eq. (\ref{4.15}) is simplified
\begin{eqnarray}
{d^{2}g_{1}\over dy^{2}}+\left[{2c+1\over y}-{1\over y-1}\right]{dg_{1}\over dy}
\nonumber\\
+\left[\,{2c-\omega^{2}/2-b\,\omega/a-b^{2}\omega^{2}/(2a^{2})\over 2\,y}+{-2c+b\omega/a\over 2\,(y-1)}\right]g_{1}=0\,
\nonumber
\end{eqnarray}

\noindent which can be identified with confluent Heun function
\begin{eqnarray}
H(\alpha,\,\beta,\,\gamma,\,\delta,\,\eta,\,z) \; ,\qquad {d^{2}H\over dz^{2}}+\left[\alpha+{1+\beta\over z}+{1+\gamma\over z-1}\right]{d H\over dz}
\nonumber\\
+\left[{1\over 2}\,{\alpha+\alpha \beta-\beta \gamma-\beta-\gamma-2\eta\over z}  +
{1\over 2}\,{\alpha \gamma+\beta+\alpha+2\eta+2\delta+\beta \gamma+\gamma\over z-1}\right]H=0
\label{4.16}
\end{eqnarray}

\noindent with parameters
\begin{eqnarray}
\alpha=0\,,
\qquad
\beta=2c\,,
\qquad
\gamma=-2\,, \qquad
\delta=-{1\over 4}\,{(a^{2}+b^{2})\,\omega^{2}\over a^{2}}\,,
\nonumber\\
\eta={1\over 4}\,{2\,ab\,\omega+(a^{2}+b^{2})\,\omega^{2}+4\,a^{2}\over a^{2}}\,,
\qquad
F_{1}=y^{\pm i\omega/2}\,H(\alpha,\,\beta,\,\gamma,\,\delta,\,\eta,\,y)\,.
\label{4.17}
\end{eqnarray}

Below we will develop a method that  makes possible to
construct solutions of the system
(\ref{4.11}) in more simple functions, solution of the Bessel equation.

\section{Additional studying of the system}

Let us perform a special transformation in  (\ref{4.11})
 (suppose   $(\alpha n - \beta m) = 1$)
\begin{eqnarray}
F_{1} =  \alpha  \; G_{1} + \beta \; G_{2} \; , \qquad
F_{2} = m \; G_{1} +  n \; G_{2} \; ;
\nonumber
\\
G_{1} =  n   \; F_{1} - \beta \; F_{2} \; , \qquad
G_{2} = - m \; F_{1} +  \alpha  \; F_{2}\; .
\label{5.1}
\end{eqnarray}

\noindent
Combining equations from  (\ref{4.11}),  we get
\begin{eqnarray}
n\; Z \; ({d \over d Z}  - a b \;  Z ) \; F_{1} - \beta \;Z \; ({d \over d  Z}  +  a b \;  Z)\; F_{2}
=
- n \; ( a^{2} Z^{2}  - \omega) \; F_{2} \; - \beta\; (b^{2}  Z^{2}- \omega ) \; F_{1} \; ,
\nonumber
\\
-m\; Z \; ({d \over d Z}  - a b \;  Z ) \; F_{1} + \alpha  \;Z \; ({d \over d  Z}  +  a b \;  Z)\; F_{2}
 =
m \; ( a^{2} Z^{2}  - \omega) \; F_{2} \; +  \alpha \; (b^{2}  Z^{2}- \omega ) \; F_{1} \; ,
\nonumber
\end{eqnarray}

\noindent from whence it follows
\begin{eqnarray}
Z \; {d \over d Z}  \; G_{1} - Z^{2} \; ab \; (nF_{1} + \beta F_{2})
=
-Z^{2} \; (na^{2} F_{2} + \beta b^{2} F_{1}) +
\omega  \; ( n F_{2} + \beta F_{1} ) \; ,
\nonumber
\\
Z \; {d \over d Z} \;  G_{2} + Z^{2} \; ab \; (m F_{1} + \alpha  F_{2})
=
Z^{2} \; ( m a^{2} F_{2} + \alpha b^{2} F_{1}) -
\omega \; ( m F_{2} + \alpha  F_{1} ) \; .
\label{5.2}
\end{eqnarray}

\noindent
Taking into account  (\ref{5.1}), eqs.  (\ref{5.2}) reduce to
\begin{eqnarray}
\left [  Z  {d \over d Z}   - Z^{2}  ab    (n    \alpha  + \beta m) +
Z^{2}     (a^{2} mn + b^{2} \alpha \beta)  -  \omega     (n  m  + \alpha \beta)   \right ]  G_{1}
\nonumber
\\
= \left [   - Z^{2}  (a   n  -  b   \beta)^{2}  + \omega   (n^{2} + \beta^{2})  \right ]   G_{2})  \; ,
\nonumber
\\[4mm]
\left [  Z  {d \over d Z}   +
 Z^{2}  ab \;  (  m \beta + n \alpha  )  -   Z^{2}  (a^{2} mn + b^{2} \alpha \beta)
 +  \omega   ( nm + \alpha \beta)  \right ]   G_{2}
\nonumber
\\
= \left [   Z^{2}  (a  m - b   \alpha)^{2}   - \omega    (m^{2} + \alpha^{2})   \right ]   G_{1}     \; .
\noindent
\label{5.3}
\end{eqnarray}

Let us impose additional restriction  (there exist two possibilities):
\begin{eqnarray}
a  n - b  \beta = 0 \qquad \Longrightarrow \qquad { \beta \over n} = {a \over b} \; ,
\nonumber
\\
\left [  Z  {d \over d Z}   - Z^{2}  ab    (n    \alpha  + \beta m) +
Z^{2}     (a^{2} mn + b^{2} \alpha \beta)  -  \omega     (n  m  + \alpha \beta)   \right ]  G_{1}
\nonumber
\\
= + \omega   (n^{2} + \beta^{2})   G_{2})  \; ,
\nonumber
\\[4mm]
\; \left [  Z  {d \over d Z}   +
 Z^{2}  ab   (  m \beta + n \alpha  )  -   Z^{2}  (a^{2} mn + b^{2} \alpha \beta)
 +  \omega   ( nm + \alpha \beta)  \right ]   G_{2}
\nonumber
\\
= \left [   Z^{2}  (a  m - b   \alpha)^{2}   - \omega \;   (m^{2} + \alpha^{2})   \right ]  G_{1}     \; .
\label{5.4}
\end{eqnarray}

\noindent
or
\begin{eqnarray}
a m - b; \alpha = 0 \qquad \Longrightarrow \qquad  {\alpha \over m} ={a \over  b} \; ,
\nonumber
\\
\left [  Z  {d \over d Z}   - Z^{2}  ab    (n    \alpha  + \beta m) +
Z^{2}     (a^{2} mn + b^{2} \alpha \beta)  -  \omega     (n  m  + \alpha \beta)   \right ]  G_{1}
\nonumber
\\
= \left [   - Z^{2}  (a   n  -  b   \beta)^{2}  + \omega   (n^{2} + \beta^{2})  \right ]   G_{2})  \; ,
\nonumber
\\[4mm]
\left [  Z  {d \over d Z}   +
 Z^{2}  ab   (  m \beta + n \alpha  )  -   Z^{2}  (a^{2} mn + b^{2} \alpha \beta)
 +  \omega   ( nm + \alpha \beta)  \right ]   G_{2}
\nonumber
\\
=  - \omega    (m^{2} + \alpha^{2})     G_{1}     \; .
\label{5.5}
\end{eqnarray}

\noindent
The two variant are equivalent each other, for definiteness
we will use the  variant (\ref{5.4}). It can be presented in more symmetrical form
\begin{eqnarray}
F_{1} =  \alpha  \; G_{1} + \beta \; G_{2}  =
+ { b \over \sqrt{a^{2} + b^{2}}}  \; G_{1} + { a \over \sqrt{a^{2} + b^{2}}}\; G_{2}
\; ,
\nonumber
\\
F_{2} = m \; G_{1} +  n \; G_{2}  =
-{ a \over \sqrt{a^{2} + b^{2}}}  \; G_{1} + { b \over \sqrt{a^{2} + b^{2}}}\; G_{2} \; ;
\label{5.6}
\end{eqnarray}

\noindent at this eqs.  (\ref{4.6})  assume the form
\begin{eqnarray}
\left [  Z {d \over d Z}   - Z^{2}  ab   {b^{2} - a^{2} \over b^{2} + a^{2} }  +
Z^{2}     ab { b^{2} - a^{2} \over b^{2} + a^{2}}  -  \omega
 ( - {ab \over a^{2} + b^{2}}+
{ab \over a^{2} + b^{2}} )   \right ]  G_{1}
\nonumber
\\
= + \omega   ( {b^{2} \over a^{2} +b^{2}}  + {a^{2} \over a^{2} +b^{2}} )   G_{2})  \; ,
\nonumber
\\[4mm]
\; \left [  Z  {d \over d Z}   +
 Z^{2}  ab   { b^{2} - a^{2}  \over a^{2} + b^{2}} -   Z^{2}  ab   { b^{2} - a^{2}  \over a^{2} + b^{2}}
 +  \omega   ( - {ab \over a^{2} + b^{2}}+
{ab \over a^{2} + b^{2}} )    \right ]   G_{2}
\nonumber
\\
= \left [   Z^{2}  ( - {a^{2} \over \sqrt{a^{2} + b^{2}}} - {b^{2} \over \sqrt{a^{2} + b^{2}}} )^{2}   -
 \omega    ( {a^{2} \over a^{2} + b^{2}} + {b^{2} \over a^{2} + b^{2}}  )  \right ]   G_{1}     \; ,
\nonumber
\end{eqnarray}

\noindent
that is
\begin{eqnarray}
 Z \; {d \over d Z}  \; G_{1}  =  \omega\;     G_{2}   \; , \qquad
 Z \; {d \over d Z}    \;  G_{2}=
    [ Z^{2} (a^{2} + b^{2})    -  \omega \;     ]  \;  G_{1}     \; .
\label{5.7}
\end{eqnarray}

From (\ref{5.7}) we derive a second order equation for  $G_{1}$
\begin{eqnarray}
\left ( Z^{2} \; {d^{2}  \over d Z^{2}} +   Z \; {d \over dZ}  +
\omega^{2} - \omega (a^{2}+b^{2}) Z^{2}  \right )
G_{1} = 0 \; .
\label{5.8}
\end{eqnarray}

To understand better the physical meaning of the equation (\ref{5.8}), it is convenient  to translate the equation to
variable  $z$, then it reads
\begin{eqnarray}
 e^{z}= \sqrt{\omega}\;  Z \,,
\qquad
\left ( {d^{2}\over dz^{2}}+\omega^{2}-(a^{2}+b^{2}) e^{2z}\right ) G_{1}=0\,.
\label{5.9}
\end{eqnarray}

It can be associated  with the Schr\"{o}dinger equation
\begin{eqnarray}
\left ( { d^{2} \over dz^{2} }   +  \epsilon - U(z)  \right )  \varphi (z) = 0
\label{5.10}
\end{eqnarray}

\noindent with potential function
$
U(z) = (a^{2}+b^{2}) e^{2z}$, and an effective force acting on the left $F_{z} = -2(a^{2}+b^{2}) e^{2z}$.
Note that when  $a=k_{1}=0, \; b=k_{2}=0$, the effective force vanishes.
The corresponding quantum-mechanical system can be illustrated by Fig.1.
\vspace{-10mm}

\begin{figure}[h!]

\unitlength=0.55mm
\begin{picture}(160,100)(-120,0)

\special{em:linewidth 0.4pt} \linethickness{0.4pt}

\put(-70,+10){\vector(+1,0){140}}  \put(+70,+5){$z$}
\put(0,-10){\vector(0,+1){80}}  \put(3,+70){$U(z)$}

\put(-60,45){\line(+1,0){120}}

\put(0,30){\circle*{1}}   \put(10,37){\circle*{1}}

\put(-9,24){\circle*{1}}

\put(-20,20){\circle*{1}}  \put(18,47){\circle*{1}}

\put(-40,15){\circle*{1}}   \put(22.5,57){\circle*{1}}  \put(25,65){\circle*{1}}

\put(-30,17){\circle*{1}}

\put(-55,13){\circle*{1}}


\put(+37,+48){$\epsilon=\omega^{2}$}

\end{picture}

\vspace{3mm}

\centering
\caption{Effective potential curve }
\end{figure}
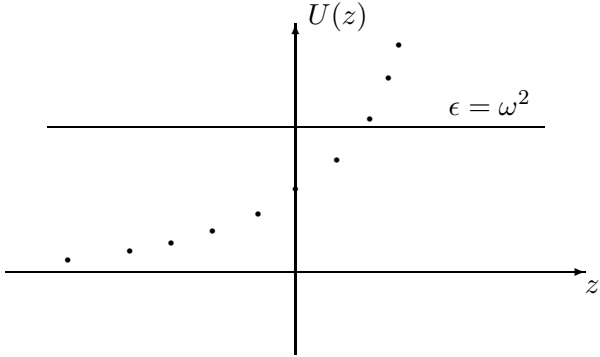


Therefore, we should expect properties of the  electromagnetic solutions
similar to those existing in  the associated  quantum-mechanical problem.

Let us turn back to eq. (\ref{5.8}) -- in the  variable
$$
x = i \; \sqrt{\omega (a^{2} + b^{2})} \; Z =  i \; \sqrt{a^{2} + b^{2}  } \; e^{z}
$$
it assumes the form of the Bessel equation
\begin{eqnarray}
\left (  {d^{2} \over dx^{2}} +{1 \over x} {d \over dx} +  1 + {\omega^{2} \over x^{2}} \right )  G_{1} = 0 \;  .
\label{5.11}
\end{eqnarray}

The first order system  (\ref{5.7}) in variable  $x$ takes the form
\begin{eqnarray}
x\,{d\over dx}\,G_{1}=\omega\,G_{2}\,, \qquad
x\,{d\over dx}\,G_{2}=-{\omega^{2}+x^{2}\over \omega}\,G_{1}\,.
\label{5.12}
\end{eqnarray}

\noindent A second order equation for $G_{2}$  reads
\begin{eqnarray}
\left[\,{d^{2}\over dx^{2}}+({1\over x}-
{2x\over \omega^{2}+x^{2}})\,{d\over dx}+{x^{2}+\omega^{2}\over x^{2}}\,\right]\,G_{2}=0 \; .
\label{5.13}
\end{eqnarray}

Note that substituting (\ref{5.6})
\begin{eqnarray}
F_{1} =   { b \over \sqrt{a^{2} + b^{2}}}   G_{1} + { a \over \sqrt{a^{2} + b^{2}}} G_{2}
\; ,\;
F_{2} =-{ a \over \sqrt{a^{2} + b^{2}}}   G_{1} + { b \over \sqrt{a^{2} + b^{2}}} G_{2}
\nonumber
\end{eqnarray}

\noindent into (\ref{4.10}), we get
\begin{eqnarray}
 f_{3} =  {e^{2z} \over \omega} ( - i b  \;   F_{1} + i a\;
 F_{2} )  = {\sqrt{a^{2} + b^{2}} \over i \; \omega} \; G_{1} \;.
\end{eqnarray}

\section{ Asymptotic  behavior of solutions
}

Mostly used  for Bessel equation  \cite{Kratzer-Franz} are solutions

\vspace{3mm}

\underline{in Bessel's functions}
\begin{eqnarray}
 G_{1}^{I} (x) = J_{+i\omega}(x)\;, \qquad
  \qquad G_{1}^{II}(x)  = J_{-i\omega}(x)\;;
\label{6.1}
\end{eqnarray}

 \underline{in Hankel's functions}
\begin{eqnarray}
 G_{1}^{I} (x) = H^{(1)}_{+i \omega}(x)\;, \qquad
 \qquad G_{1}^{II} (x) = H^{(2)}_{+i \omega}(x)\;,
\nonumber
\\
I'  \qquad G_{1}(x)  = H^{(1)}_{-i\omega}(x)\;, \qquad
II'  \qquad G_{1}(x)  = H^{(2)}_{-i\omega}(x)\; ;
\label{6.2}
\end{eqnarray}

\noindent
note that $H^{(1)}_{-i\omega}(x) = e^{-\omega \pi} H^{(2)}_{i\omega}(x)$),
so the primed cases $I', II'$ coincide respectively with  $II,
I$  and by this reason wil not be considered below;

\underline{in Neyman functions}
\begin{eqnarray}
G_{1}^{I} (x) = N_{+i\omega}(x)\;, \qquad
 G_{1}^{II}(x)  = N_{-i\omega}(x)\; .
\label{6.3}
\end{eqnarray}

For shortness, below the notation $
+ \sqrt{a^{2} + b^{2}} = 2 \sigma$ is used.
First, let us consider solutions in Bessel's functions \cite{Kratzer-Franz}  when

\vspace{2mm}
$
z \rightarrow - \infty,\; x  = i \sigma e^{z} \rightarrow i 0\;,
$
\begin{eqnarray}
 G_{1}^{I} (x) = J_{+i\omega}(x) = {1 \over \Gamma(1+ i \omega )} \; ({x \over 2})^{+i\omega} =
 {( i \; \sigma )^{i\omega} \over \Gamma(1+ i \omega )} \;
 \; e^{+i\omega z}\,   ,
\nonumber
\\
 G_{1}^{II}(x)  = J_{-i \omega}(x) = {1 \over \Gamma(1 - i \omega )} \; ({x \over 2})^{-i\omega}  =
 {(i \; \sigma )^{-i\omega}  \over \Gamma(1 - i \omega )} \;  \; e^{-i\omega z} \,.
\label{6.4}
\end{eqnarray}

In the region
$
z \rightarrow + \infty,\; (x = i \sigma e^{z} = i X \rightarrow   i \infty  )$,
using the knows asymptotic formula \cite{Kratzer-Franz}
\begin{eqnarray}
J_{i \omega}(x) \sim \sqrt{{2 \over \pi x}} \; \cos \left  ( x - (i\omega +{1\over 2}) {\pi\over 2} \right ) ,
\nonumber
\end{eqnarray}

\noindent
we get
\begin{eqnarray}
 G_{1}^{I} (z \rightarrow \infty )  = J_{+i\omega}(z \rightarrow \infty ) \sim
 e^{i\pi /4} \;  \sqrt{{1 \over 2 \pi i  X}} \;e^{-\omega \pi /2}   \;  e^{+X}
 \; ,
\nonumber
\\
 G_{1}^{II} (z \rightarrow \infty )=J_{-i\omega}(z \rightarrow \infty )  \sim
 e^{i\pi /4} \;  \sqrt{{1 \over 2 \pi i  X}} \;e^{+\omega \pi /2}   \;  e^{+X}
 \; .
\label{6.5}
\end{eqnarray}

Let us consider solutions in Hankel's functions \cite{Kratzer-Franz}, determined   in terms of
 $J_{\pm i\omega}(x)$ as follows
 \begin{eqnarray}
H^{(1)}_{i\omega} (x)  = +
{i \over \sin (i\omega \pi) } \left ( e^{\omega \pi} J_{+i\omega}(x) -  J_{-i\omega}(x) \right  )
\; ,
\nonumber\\
H^{(2)}_{i\omega} (x) =- {i \over \sin (i\omega \pi)  } \left ( e^{-\omega \pi} J_{+i\omega}(x) -  J_{-i\omega}(x) \right  )\; .
\label{6.6}
\end{eqnarray}

\noindent so that
$
z \rightarrow - \infty,\; x \rightarrow i 0\;,
$
\begin{eqnarray}
G_{1}^{I} (x) = H^{(1)}_{i\omega} (x)  =
+ {i \over \sin (i\omega \pi) }\left ( e^{+\omega \pi}  {( i \; \sigma )^{i\omega} \over \Gamma(1+ i \omega )}
 \; e^{+i\omega z}
-
 {( i \; \sigma )^{-i\omega} \over \Gamma(1- i \omega )} \; e^{-i\omega z} \right  )
\; ,
\nonumber
\\
G_{1}^{II} (x) = H^{(2)}_{i \omega} (x) =-
{i \over \sin (i\omega  \pi) } \left (  e^{-\omega  \pi}
{( i \; \sigma )^{i\omega} \over \Gamma(1+ i \omega )}  \; e^{+i\omega z}
-  {( i \; \sigma )^{-i\omega} \over \Gamma(1- i \omega )}  \; e^{+i\omega z}  \right  )\; .
\nonumber
\\
\label{6.7}
\end{eqnarray}

Behavior of them  when   $z \rightarrow + \infty$ is governed the  known relation \cite{Kratzer-Franz}
\begin{eqnarray}
H^{(1)}_{i\omega} (x) \sim \sqrt{{2 \over \pi x}} \; \exp \left [ + i \left
(x - {\pi \over 2} (i\omega +{1\over 2}) \right ) \right ] ,
\nonumber
\\
H^{(2)}_{i\omega} (x) \sim \sqrt{{2 \over \pi x}} \; \exp
\left [ -i \left (x - {\pi \over 2} (i\omega +{1\over 2}) \right ) \right ] ;
\nonumber
\end{eqnarray}

\noindent
from whence it follows

\vspace{2mm}
$
z \rightarrow + \infty,\; x  =  i X\rightarrow    i\infty  \;,
$
\begin{eqnarray}
G_{1}^{I} (x) =  H^{(1)}_{i \omega} (x) \sim
  e^{-i\pi /4} \sqrt{{2 \over i \pi X}} \; e^{+ \omega \pi /2 } \; e^{-X} \; ,
\nonumber
\\
G_{1}^{II} (x) = H^{(2)}_{i \omega} (x) \sim
e^{+i\pi /4} \sqrt{{2 \over i \pi X}} \; e^{ -\omega \pi /2 } \; e^{+X}
 \; .
\label{6.8}
\end{eqnarray}

Let us consider  interpretation of the first type solution:
this wave goes from the left, then it  is partly  reflected and partly goes forward through an
 effective potential barrier but gradually  damping as $z$ rises.
 The corresponding reflection coefficient is determined as  follows
  \begin{eqnarray}
G(z) \sim M_{+}^{I} e^{+i\omega z} + M_{-}^{I} e^{-i \omega z} \;, \qquad R = { \mid M_{-} ^{I}\mid^{2} \over
\mid M_{+}^{I}  \mid^{2} }  \; .
\label{6.9}
\end{eqnarray}

\noindent Taking into account identities
\begin{eqnarray}
(i\sigma)^{+i\omega} = (e^{i\pi/2} e^{\ln \sigma } )^{+i\omega} =e^{-\omega \pi /2}  e^{+i \omega \ln \sigma } \; ,
\nonumber
\\
(i\sigma)^{-i\omega} = (e^{i\pi/2} e^{\ln \sigma } )^{-i\omega} =e^{+\omega \pi /2}  e^{-i \omega \ln \sigma } \; ;
\label{6.10}
\end{eqnarray}

\noindent  we derive
\begin{eqnarray}
\mid M_{+}^{I}  \mid^{2}  = {1 \over  \sin (+i\omega \pi)  \sin (-i\omega \pi) }\;
 {e^{+\omega \pi}  \over \Gamma(1- i \omega ) \Gamma(1+ i \omega ) }\; ,
\nonumber
\\
\mid M_{-}^{I}  \mid^{2}  = {1 \over  \sin (+i\omega \pi)  \sin (-i\omega \pi) }\;
 {e^{+\omega \pi}  \over \Gamma(1- i \omega ) \Gamma(1+ i \omega ) }\; .
\label{6.11}
\end{eqnarray}

\noindent
This means that for all solutions of that type the reflection coefficient always equals to $1$:
\begin{eqnarray}
R= 1\;.
\label{6.12}
\end{eqnarray}

Solutions of the second type, rising to infinity as $z \rightarrow + \infty$,
are characterized by
\begin{eqnarray}
M_{+}^{II} e^{+i\omega z} + M_{-}^{II} e^{-i \omega z} \;, \qquad R = { \mid M_{-}^{II} \mid^{2} \over
\mid M_{+} ^{II} \mid^{2} }  = e^{4\omega \pi } > 1  \; .
\label{6.13}
\end{eqnarray}

Finally, let us specify asymptotic behavior of solutions in terms of Neyman functions.
They functions are defined by \cite{Kratzer-Franz}
\begin{eqnarray}
N_{i\omega}(x) = {  \cos (i\omega\pi) \; J_{i\omega} (x) - J_{-i\omega}(x)  \over \sin (i\omega \pi) } \; ,
\nonumber
\\
N_{-i\omega}(x) = {  J_{i\omega} (x) - \cos (i\omega \pi) \;  J_{-i\omega}(x)  \over \sin (i\omega \pi) } \; .
\label{6.14}
\end{eqnarray}

In the region
$
z \rightarrow + \infty,\; (x  = i X \rightarrow  i\infty  )$,
with  the use of the known relation \cite{Kratzer-Franz}
\begin{eqnarray}
N_{i\omega}(x) \sim
 \sqrt{{2 \over i \pi X}} \; \sin \left  ( i X - (i \omega +{1\over 2}) {\pi\over 2} \right )  ,
 \nonumber
 \end{eqnarray}

\noindent
we get
\begin{eqnarray}
G_{1}^{I} (x) = N_{+i \omega(x)} \sim
i e^{+i\pi/4}\sqrt{{1 \over  2 i \pi   X}}  e^{- \omega\pi  /2} \; e^{X} \; ,
\nonumber
\\
 G_{1}^{II}(x)  = N_{-i \omega}(x) \sim
+i e^{+i\pi/4}\sqrt{{1 \over 2 i \pi   X}}  e^{+ \omega\pi  /2} \; e^{X}
 \;.
\label{6.15}
\end{eqnarray}

In the region  $z\rightarrow -\infty$  their behavior is given by
\begin{eqnarray}
G^{I}(z) =  {  \cos (i\omega\pi) \over \sin (i\omega \pi) } \; {( i \; \sigma )^{i\omega} \over \Gamma(1+ i \omega )} \;
 \; e^{+i\omega z}- {1 \over \sin (i\omega \pi) } {(i \; \sigma )^{-i\omega}  \over \Gamma(1 - i \omega )} \;  \; e^{-i\omega z} \; ,
\nonumber
\\
G^{II}(z) = {  1 \over \sin (i\omega \pi) } \,{( i \; \sigma )^{i\omega} \over \Gamma(1+ i \omega )} \;
 \; e^{+i\omega z}-{ \cos (i\omega \pi)  \over \sin (i\omega \pi) }  \; {(i \; \sigma )^{-i\omega}  \over \Gamma(1 - i \omega )} \;  \; e^{-i\omega z}\; .
\label{6.16}
\end{eqnarray}

\noindent For these solutions we have respectively
\begin{eqnarray}
R^{I} =  {e^{2\omega \pi} \over  (e^{2\omega \pi} + e^{-2\omega \pi} )/4}=
{4 \over  1 + e^{-4\omega \pi} }\; ,
\nonumber
\\
R^{II} = e^{2\omega \pi} \; (e^{2\omega \pi} + e^{-2\omega \pi})/4 \; = {1 + e^{4\omega \pi} \over 4}\;.
\label{6.17}
\end{eqnarray}

\section{On explicit form of the function $G_{2}$ }

 The function $G_{1}(x)$ satisfies the Bessel equation
\begin{eqnarray}
\left (  {d^{2} \over dx^{2}} +{1 \over x} {d \over dx} +  1 + {\omega^{2} \over x^{2}} \right )  G_{1} = 0 \;  ;
\label{7.1}
\end{eqnarray}

\noindent
the second function $G_{2}(x)$ is determined by
\begin{eqnarray}
G_{2} = { x \over  \omega } \,{d\over dx}\,G_{1} \; .
\label{7.2}
\end{eqnarray}

\noindent
Solutions of the Bessel equation obey
the following recurrent formulas  \cite{Kratzer-Franz}
 \begin{eqnarray}
  x {d \over dx }  \; F _{ i \omega}  =   i \omega  \;  F _{i \omega}    - x F_{i \omega  +1} \; ,
\nonumber
\\
 x {d \over  dx } \; F _{-i \omega}  = + i \omega  \;  F _{-i \omega} (x)  + x F_{-i \omega -1  } \; ,
\label{7.3}
\end{eqnarray}

\noindent
where $F_{\pm \nu}$  stands for
\begin{eqnarray}
J_{\pm \nu}(x)\;, \qquad
  H^{(1)}_{\pm \nu}(x)\;, \qquad  H^{(2)}_{\pm \nu}(x)\;, \qquad  N_{\pm \nu}(x)\; .
  \nonumber
  \end{eqnarray}

Therefore, with the help of (\ref{7.3}), one can express $G_{2}$ in terms of the known
$G_{1}$. For instance,
\begin{eqnarray}
G_{1}^{I} (x)  = H^{(1)}_{+i \omega} (x) \;, \qquad
G_{2}^{I} (x)  =
 i   \;  H^{(1)} _{+i\omega} (x)    - { x  \over \omega }\; H^{(1)}_{i \omega  +1} (x)  \; ,
\nonumber
\\
G_{1}^{II} (x) = H^{(2)}_{+i \omega} (x) \;, \qquad
G_{2}^{II} =
 i  \;  H^{(2)} _{+i\omega} (x)    - {x \over \omega }\;  H^{(2)}_{i \omega  +1}(x) \; .
\label{7.4}
\end{eqnarray}

\noindent
Remember that
\begin{eqnarray}
F_{1}^{I} =   { b \over \sqrt{a^{2} + b^{2}}}   G_{1} + { a \over \sqrt{a^{2} + b^{2}}} G_{2}
\; ,\nonumber
\\
F_{2}^{I} =-{ a \over \sqrt{a^{2} + b^{2}}}   G_{1} + { b \over \sqrt{a^{2} + b^{2}}} G_{2} \; ,
\nonumber
\\
 f_{3}^{I} =  {e^{2z} \over \omega} ( - i b  \;   F_{1}^{I} + i a\;
 F_{2}^{I} )  = {\sqrt{a^{2} + b^{2}} \over i \; \omega} \; G_{1} \;.
\label{7.5}
\end{eqnarray}

Let us examine  asymptotic behavior of $G_{2}$.
Starting with
\begin{eqnarray}
H^{(1)}_{i\omega} (x)  = + {i \over \sin (i\omega \pi) } \left ( e^{\omega \pi} J_{+i\omega}(x) -  J_{-i\omega}(x) \right  )
\; ,
\nonumber
\\
H^{(2)}_{i\omega} (x) =- {i \over \sin (i\omega \pi)  } \left ( e^{-\omega \pi} J_{+i\omega}(x) -  J_{-i\omega}(x) \right  )\; ,
\nonumber
\\
H^{(1)}_{i\omega+1} (x)  = + {i \over \sin (i\omega+1) \pi } \left ( e^{-i(i\omega+1)\pi } J_{+i\omega+1}(x) -  J_{-(i\omega+1)}(x) \right  )
\; ,
\nonumber
\\
H^{(2)}_{i\omega+1} (x) =- {i \over \sin (i\omega+1) \pi  }
\left (  e^{i(i\omega+1)\pi} J_{+i\omega+1}(x) -  J_{-(i\omega+1)}(x) \right  )\; ,
\label{7.6}
\end{eqnarray}

\noindent with the help of relations

\vspace{2mm}

$
z \rightarrow - \infty,\; x \rightarrow i 0\;,
$
\begin{eqnarray}
 J_{+i\omega}(x) \sim  {( i \; \sigma )^{i\omega} \over \Gamma(1+ i \omega )} \;
 \; e^{+i\omega z}\,   ,\qquad
 J_{-i \omega}(x) \sim
 {(i \; \sigma )^{-i\omega}  \over \Gamma(1 - i \omega )} \;  \; e^{-i\omega z} \,.
\nonumber
\end{eqnarray}

\noindent we get

$
z \rightarrow - \infty,\; x \rightarrow i 0\;,
$
\begin{eqnarray}
 H^{(1)}_{i\omega}   \sim
+ {i \over \sin (i\omega \pi) }\left ( e^{+\omega \pi}  {( i \; \sigma )^{i\omega} \over \Gamma(1+ i \omega )} \; e^{+i\omega z}
-
 {( i \; \sigma )^{-i\omega} \over \Gamma(1- i \omega )} \; e^{-i\omega z} \right  )
\; ,
\nonumber
\\
 H^{(2)}_{i \omega}  \sim -
{i \over \sin (i\omega  \pi) } \left (  e^{-\omega  \pi}
{( i \; \sigma )^{i\omega} \over \Gamma(1+ i \omega )}  \; e^{+i\omega z}
-  {( i \; \sigma )^{-i\omega} \over \Gamma(1- i \omega )}  \; e^{-i\omega z}  \right  )\;,
\nonumber
\end{eqnarray}

\begin{eqnarray}
H^{(1)}_{i\omega+1}   \sim  {i \over \sin (i \omega+1) \pi }\left ( e^{-i(i\omega+1)\pi}
{(i \sigma )^{i\omega +1} \over \Gamma(2+ i \omega  )}
 e^{i \omega z}e^{z}   -
{(i \sigma )^{-i\omega-1} \over \Gamma(- i \omega   )} e^{-i\omega z} e^{-z}   \right  )
\nonumber
\\
\sim -{i \over \sin (i \omega+1) \pi }\,{(i \sigma )^{-i\omega-1}\over \Gamma(- i \omega  )}  \;e^{-i\omega z} e^{-z}
\; ,
\nonumber
\\
\nonumber
\end{eqnarray}

\begin{eqnarray}
H^{(2)}_{i\omega+1} (x) \sim
   {-i \over \sin (i \omega+1)\pi } \left (  e^{i(i \omega +1) \pi} {(i \sigma  )^{i \omega +1}
\over \Gamma(2+ i \omega  )} \; e^{i\omega z} e^{z }
-  {(i \sigma  )^{-i \omega-1} \over \Gamma(- i \omega  )} e^{-i\omega z} e^{-z }  \right  )
\nonumber
\\
\sim  {i \over \sin (i \omega+1)\pi }\, {(i \sigma  )^{-i \omega-1}\over \Gamma(- i \omega  )}  \; e^{-i\omega z} e^{-z }
\; .
\nonumber
\\
\label{7.7}
\end{eqnarray}

So we get
\begin{eqnarray}
G_{2}^{I} (x)  =
  -{1 \over \sin (i\omega \pi) }\left ( e^{+\omega \pi}  {( i \; \sigma )^{i\omega} \over \Gamma(1+ i \omega )} \; e^{+i\omega z}
-
 {( i \; \sigma )^{-i\omega} \over \Gamma(1- i \omega )} \; e^{-i\omega z} \right  )
\nonumber
\\
 - { 2 \sigma    \over \omega }\,{1 \over \sin (i \omega+1) \pi }\,
{(i \sigma )^{-i\omega-1}\over \Gamma(- i \omega  )}   \;e^{-i\omega z}
  \; .
\label{7.8}
\end{eqnarray}

Taking into consideration  an identity
\begin{eqnarray}
- { 2 \sigma    \over \omega }\,{1 \over \sin (i \omega+1) \pi }\,
{(i \sigma )^{-i\omega-1}\over \Gamma(- i \omega  )}   \;e^{-i\omega z}
\nonumber
\\
=
+{ 2 \sigma    \over \omega }\,{1 \over \sin (i \omega \pi) }
{(i \sigma )^{-i\omega} (-i\omega) \over (i \sigma ) \Gamma(1 - i \omega  )}   \;e^{-i\omega z}=
- 2 \,{1 \over \sin (i \omega \pi) }
{(i \sigma )^{-i\omega}  \over  \Gamma(1 - i \omega  )}   \;e^{-i\omega z}
\label{7.9}
\end{eqnarray}

\noindent one reduces the above relation (\ref{7.8}) to the form
\begin{eqnarray}
G_{2}^{I} (x)  =
  -{1 \over \sin (i\omega \pi) }\left ( e^{+\omega \pi}  {( i \; \sigma )^{i\omega}
   \over \Gamma(1+ i \omega )} \; e^{+i\omega z}
+
 {( i \; \sigma )^{-i\omega} \over \Gamma(1- i \omega )} \; e^{-i\omega z} \right  ).
\label{7.10}
\end{eqnarray}

\noindent
In similar manner one  can treat the case
\begin{eqnarray}
G_{2}^{II} =
{1 \over \sin (i\omega  \pi) } \left (  e^{-\omega  \pi}
{( i \; \sigma )^{i\omega} \over \Gamma(1+ i \omega )}  \; e^{+i\omega z}
-  {( i \; \sigma )^{-i\omega} \over \Gamma(1- i \omega )}  \; e^{-i\omega z}  \right  )
\nonumber
\\
 + {2 \sigma  \over \omega }\,{1 \over \sin (i \omega+1)\pi }\,
{(i \sigma  )^{-i \omega-1}  \over \Gamma(- i \omega  )} \; e^{-i\omega z}
 \nonumber
 \\
=  {1 \over \sin (i\omega  \pi) } \left (  e^{-\omega  \pi}
{( i \; \sigma )^{i\omega} \over \Gamma(1+ i \omega )}  \; e^{+i\omega z}
+  {( i \; \sigma )^{-i\omega} \over \Gamma(1- i \omega )}  \; e^{-i\omega z}  \right  )\; .
\label{7.11}
\end{eqnarray}

Behavior of these solutions when   $z \rightarrow + \infty$ is governed the   relation
\begin{eqnarray}
H^{(1)}_{i\omega} (x) \sim \sqrt{{2 \over \pi x}} \; \exp \left [ + i \left
(x - {\pi \over 2} (i\omega +{1\over 2}) \right ) \right ] ,
\nonumber
\\
H^{(2)}_{i\omega} (x) \sim \sqrt{{2 \over \pi x}} \; \exp
\left [ -i \left (x - {\pi \over 2} (i\omega +{1\over 2}) \right ) \right ] ;
\nonumber
\end{eqnarray}

\noindent
from whence it follows
\begin{eqnarray}
 H^{(1)}_{i \omega} (x) \sim
  e^{-i\pi /4} \sqrt{{2 \over i \pi X}} \; e^{+ \omega \pi /2 } \; e^{-X} \; ,
\nonumber
\\
H^{(2)}_{i \omega} (x) \sim
e^{+i\pi /4} \sqrt{{2 \over i \pi X}} \; e^{ -\omega \pi /2 } \; e^{+X}
\nonumber
\\
H^{(1)}_{i \omega +1} (x) \sim
 \sqrt{{2 \over i \pi X}} \; \exp \left [ + i \left
(i X - {\pi \over 2} (i \omega +1 +{1\over 2}) \right ) \right ]
\nonumber
\\
\sim
 -i\;e^{-i \pi/4}
 \sqrt{{2 \over i \pi X}} \; e^{ +\omega \pi/2 } \; e^{-X} \; ,
\nonumber
\\
H^{(2)}_{i \omega +1} (x) \sim
 \sqrt{{2 \over i \pi X}} \; \exp \left [ -i \left
 (iX - {\pi \over 2} (i \omega +1  +{1\over 2}) \right ) \right ]
 \nonumber
 \\
 \sim i\;e^{+i \pi/4}
 \sqrt{{2 \over i \pi X}} \; e^{ -\omega \pi/2 } \; e^{+X} \; .
\label{7.12}
\end{eqnarray}

Therefore, we arrive at the formulas
\begin{eqnarray}
G_{2}^{I} (x)  =
 i   \;  H^{(1)} _{i \omega} (x)    - { x  \over \omega }\; H^{(1)}_{i \omega  +1} (x)
\nonumber
\\
\sim
i  e^{-i\pi /4} \sqrt{{2 \over i \pi X}} \; e^{+ \omega \pi /2 } \; e^{-X}- {  X  \over \omega }\,e^{-i \pi/4}
 \sqrt{{2 \over i \pi X}} \; e^{ +\omega \pi/2 } \; e^{-X}\; ,
\nonumber
\\
G_{2}^{II} =
 i  \;  H^{(2)} _{i \omega} (x)    - {x \over \omega }\;  H^{(2)}_{i \omega  +1}(x)
\nonumber
\\
 \sim
 i e^{+i\pi /4} \sqrt{{2 \over i \pi X}} \; e^{ -\omega \pi /2 } \; e^{+X} + {X \over \omega }\;e^{+i \pi/4}
 \sqrt{{2 \over i \pi X}} \; e^{ -\omega \pi/2 } \; e^{+X}\; .
\label{7.13}
\end{eqnarray}

Evidently, to find asymptotic for  $G_{2}$,  it is sufficient to  make use of the known asymptotic
for $G_{1}$. For instance,
\begin{eqnarray}
G_{2}^{I}  \sim  { 1 \over  \omega } \,{d\over dz}\,
 {i \over \sin (i\omega \pi) }\left ( e^{+\omega \pi}  {( i \; \sigma )^{i\omega} \over \Gamma(1+ i \omega )}
 \; e^{+i\omega z}
-
 {( i \; \sigma )^{-i\omega} \over \Gamma(1- i \omega )} \; e^{-i\omega z} \right  )
\nonumber
\\
= -
{1 \over \sin (i\omega \pi) }\left ( e^{+\omega \pi}  {( i \; \sigma )^{i\omega} \over \Gamma(1+ i \omega )}
 \; e^{+i\omega z}
+
 {( i \; \sigma )^{-i\omega} \over \Gamma(1- i \omega )} \; e^{-i\omega z} \right  )\;;
\label{7.14}
\end{eqnarray}

\noindent which coincides with (\ref{7.10}).
It is a superposition of two plane waves with reflection coefficient $R=1$.

\section{Concluding remarks }

In accordance with (\ref{5.10}), an equation  below
\begin{eqnarray}
\omega^{2} = U (z) \qquad \omega^{2} = (a^{2} + b^{2}) e^{2z_{0}}
\label{10.1}
\end{eqnarray}

\noindent determines a critical  point $z_{0}$ in which behavior  of the function $G_{1}(x)$
must change dramatically. To such a point   $z_{0}$ there corresponds
\begin{eqnarray}
x_{0} = i \sqrt{a^{2} + b^{2}} e^{z_{0}} =  i \omega \; .
\label{10.2}
\end{eqnarray}

In order to examine behavior of solutions in vicinity of  $x_{0}$,
it is convenient to introduce a new coordinate
\begin{eqnarray}
x =x_{0} + i\omega \; u =  i \omega ( 1 +  u ) \;, \qquad {d \over dx} = {1 \over  i \omega } \; {d \over d u} \; ;
\label{10.3}
\end{eqnarray}

\noindent eq. (\ref{5.11}) for  $G_{1}(x)$ assumes the form
\begin{eqnarray}
\left (  {d^{2} \over du^{2}} + {1 \over 1 + u}
{d \over du}  -\omega^{2} +  {\omega^{2}   \over (1+ u)^{2} } \right )  G_{1} = 0 \;  .
\label{10.4}
\end{eqnarray}

Close to  $u=0$,  we have
\begin{eqnarray}
\left (  {d^{2} \over du^{2}} +
{d \over du}  \right )  G_{1} = 0 \;  .
\end{eqnarray}
that is
$$
G_{1} = e^{Bu} , \qquad B^{2} + B = 0,  \qquad  B =0, -1 \; ;
\label{10.5}
$$

\noindent physically interesting is the choice $B=-1$.

To such a critical value $x_{0} = i \omega$, there correspond
\begin{eqnarray}
\omega = \sqrt{k_{1}^{2} +k_{2}^{2} } \; e^{z_{0}} \qquad \Longrightarrow \qquad
z_{0} = \ln { \omega \over \sqrt{k_{1}^{2} +k_{2}^{2} } } \; ;
\label{10.6}
\end{eqnarray}

\noindent
in usual units, this relation reads

\begin{eqnarray}
z_{0} =  \rho \; \ln { \omega  \over c\; \sqrt{k_{1}^{2} +k_{2}^{2} } } \; ,
\label{10.6}
\end{eqnarray}

\noindent where $\rho$ is a curvature radius of the Lobachevsky space.

\vspace{10mm}

Let us summarize results.

\begin{quotation}

Lobachevsky geometry simulates a  medium with special constitutive relations.
The situation is specified in quasi-cartesian coordinates $(x,y,z)$.
Exact solutions of the Maxwell equations in  complex  3-vector ${\bf E}+i {\bf B}$ form, extended to curved
space models within the tetrad formalism, have been found in Lobachevsky space.
The problem reduces to a second order diffe\-rential equation which can be associated with an 1-dimensional
Schr\"{o}dinger problem for a particle in external  potential field $U(z) =  U_{0}e^{2z}$.

In quantum mechanics, curved geometry acts as an effective potential barrier with reflection coefficient $R=1$;
in electrodynamic  context  results similar to quan\-tum-mechanical  ones   arise:
the Lobachevsky  geometry   simu\-lates  a  medium  that effectively  acts as
an ideal  mirror. Penetration of the electromagnetic field into the effective medium, depends on the parameters of an
electromagnetic wave,
frequency $\omega, \; k_{1}^{2} + k_{2}^{2}$, and the curvature radius $\rho$ -- see (\ref{10.6}).
See  illustrations in Fig. 2,3.

\begin{figure}[h!]
\centering
\includegraphics{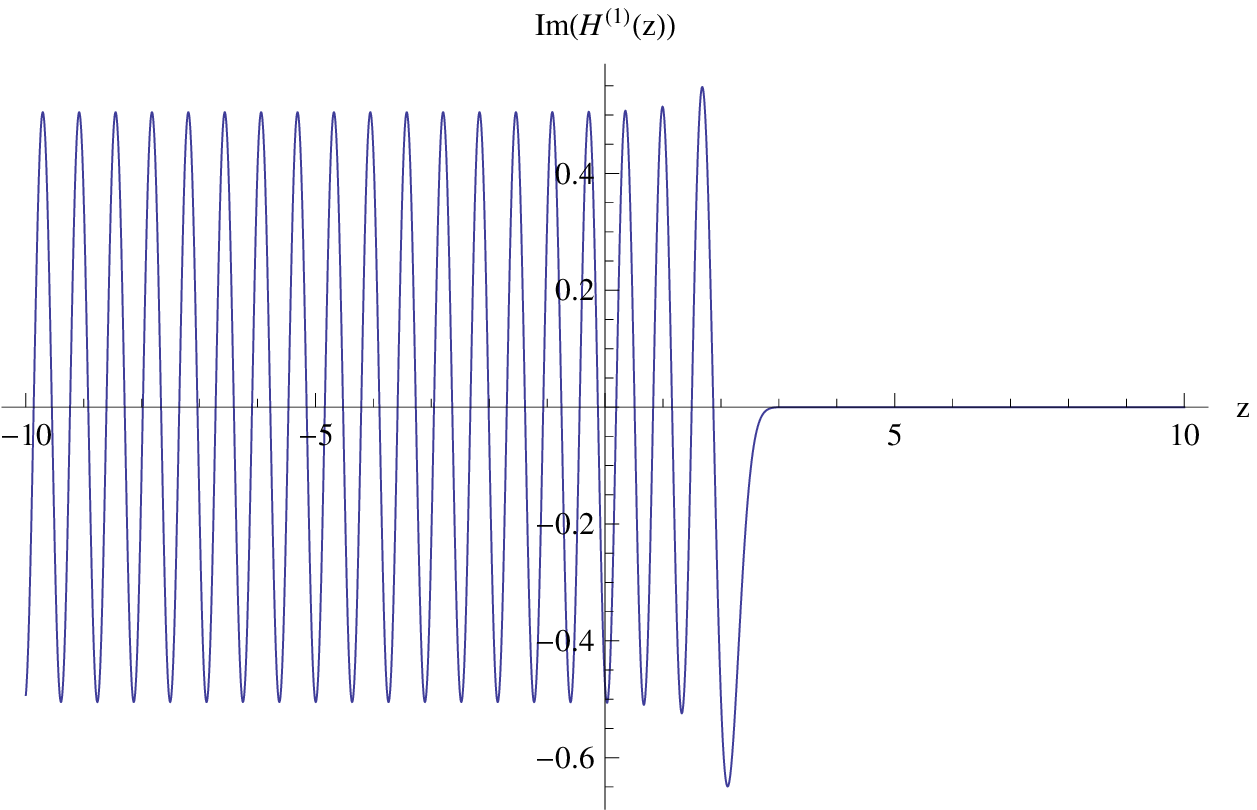} \caption{$\mbox{Im}\;H^{(1)}_{+i\omega }, \; \omega = 10$}
\end{figure}

\begin{figure}[h!]
\centering
\includegraphics{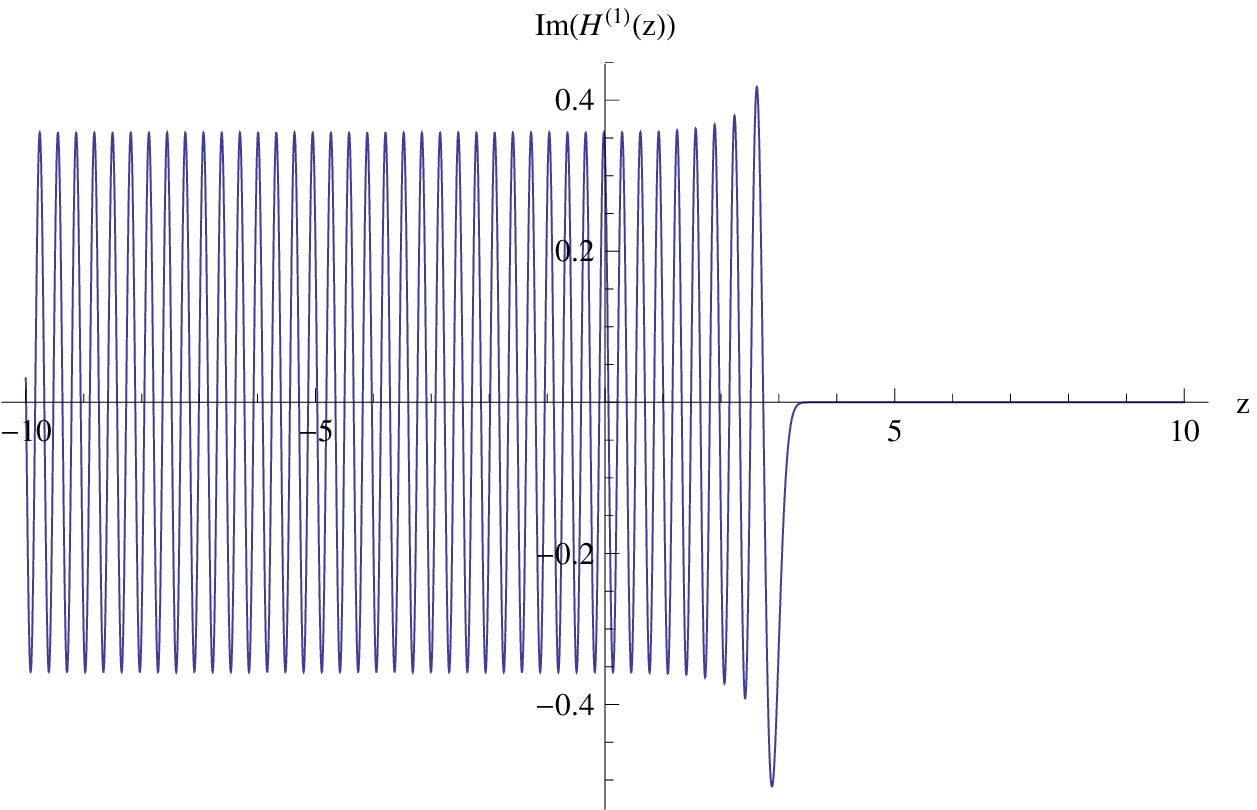} \caption{$\mbox{Im}\;H^{(1)}_{+i\omega }, \; \omega = 20$}
\end{figure}

\end{quotation}

\newpage

\section{Acknowledgement}

Authors are grateful to Dr. G.G. Krylov for help and advices.
This  work was   supported   by the Fund for Basic Researches of Belarus,
 Grant F11M-152.


\begin{thebibliography}{99}



\bibitem{Olevsky}
M.N. Olevsky. Three-orthogonal  coordinate systems in spaces of constant curvature,
in which equation  $\Delta_{2}U + \lambda
U=0$ permits the  full separation of variables.
Mathematical collection.  1950. Vol.  27. P. 379 -- 426.




\bibitem{1901-Weber}
Weber H.   Die partiellen
Differential-Gleichungen der mathematischen Physik nach Riemann's
Vorlesungen.
 Friedrich Vieweg und Sohn.  Braunschweig. 1901. P. 348.




\bibitem{1907-Silberstein(1)}
Silberstein L.
Elektromagnetische Grundgleichungen in bivectorieller Behandlung.
//  Ann.   Phys. 1907. Bd.  22.  S.   579 -- 586.



\bibitem{1931-Oppenheimer}
 Oppenheimer J.  Note on light Quanta and the electromagnetic
field //  Phys. Rev.  1931. Vol.  38. P.    725 -- 746.




\bibitem{1931-Majorana}
Majorana E.  Scientific Papers.  Unpublished. Deposited at
the "Domus Galileana". Pisa, quaderno 2, p. 101/1; 3, p. 11, 160;
15, p. 16; 17, p. 83, 159.







\bibitem{1914-Marcolongo}
 Marcolongo R.
 Les transformations de Lorentz et les \'equations
de l'\'electrodynamique // Annales de la Facult\'{e} des Sciences de Toulouse.
1914. Vol.  4. P.  429 -- 468.



\bibitem{1915-Bateman}
Bateman H.,
The Mathematical analysis of electrical and
Optical wave-Motion on the basis of Maxwells equations. Cambridge
University Press, 1915.






\bibitem{1941-Tonnelat}
 Tonnelat  M.  Sur la th\'{e}orie du photon dans un espace de Riemann //
 Ann. Phys. N.Y. 1941. Vol. 15. P. 144.



\bibitem{1956-Borgardt}
Borgardt. Wave equations for a phiton.  // JETP. 158.    {\bf 34}
(1958) 1323-1325.



\bibitem{1957-Kuohsien}
 Kuohsien T.  Sur les theories matricielles du photon //
 C.  R
Acad. Sci. Paris.  1857. Vol.  245. P.   141 -- 144.






\bibitem{1958-Lomont}
 Lomont J. Dirac-like wave equations for particles of zero rest
mass and their quantization //  Phys. Rev.  1958. Vol.  11. P. 1710 -- 1716.






\bibitem{1962-Sachs-Schwebel}
 Sachs M.,  Schwebel S. On covariant formulations of
the Maxwell-Lorentz theory of electromagnetism //
 J. Math. Phys. 1962. Vol. 3. P.   843 -- 848.


\bibitem{1964-Ellis}
 Ellis  J.  Maxwell's equations and theories of Maxwell form //
Ph.D. thesis. University of London.  1964.  417 p.



\bibitem{1974-Recami1}
 Mignani R.,  Recami E.,  Baldo  M.,
  About a Dirac-like equation for
the photon, according to E. Majorana //  Lett. Nuovo Cimento.  1974. Vol. 11. P. 568 -- 572.





\bibitem{1975-Edmonds}
 Edmonds J.
   Comment on the Dirac-like equation for the
photon // Nuovo Cim. Lett.   1975. Vol.  13. P.  185 -- 186.






\bibitem{1980-Silveira}
 Da Silveira A.
Invariance algebras  of the Dirac and Maxwell
equations // Nouvo Cim.  A.  1980. Vol.  56. P. 385 -- 395.







\bibitem{1981-Chow}
 Chow T.  A Dirac-like equation for the photon
 //  J. Phys. A.  1981. Vol. 14. P.  2173 -- 2174.



\bibitem{1983-Fushchich}
 Fushchich V.I.   Nikitin A.G.  Symmetries of Maxwell's equations.
   Kluwer. Dordrecht. 1987.



 \bibitem{1982-Cook(1)}
 Cook R.  Photon dynamics //
 Phys. Rev. A.  1982. Vol. 25. P.  2164 -- 2167;
    Lorentz covariance of photon dynamics
//  Phys. Rev. A.  1982. Vol. 26. P.  2754 -- 2760.








\bibitem{1990-Recami}
 Recami E.
  Possible physical meaning of the photon wave-function,
according to Ettore Majorana // Hadronic Mechanics and
Non-Potential Interactions.  New York, 1990.   P. 231 -- 238.





\bibitem{1990-Inagaki}
 Inagaki T.
  Quantum-mechanical approach to a free photon //
    Phys. Rev. A.    1994. Vol.  49. P.  2839 -- 2843.





\bibitem{1994-Bialynicki-Birula}
 Bialynicki-Birula I.
  On the wave function of the photon // Acta
Phys. Polon.   1994. Vol.  86. P.   97 -- 116;
Photon wave function //   Progress in
Optics. 1996. Vol. 36. P.  248 -- 294; arXiv:quant-ph/050820.


\bibitem{2005-Birula}
 Bialynicki-Birula I.,   Bialynicka-Birula Z.  Beams of
electromagnetic radiation carrying angular momentum: The
Riemann -- Silberstein vector and the classical-quantum
correspondence //  arXiv:quant-ph/0511011.



\bibitem{1995-Sipe}
 Sipe J.   Photon wave functions//
 Phys. Rev. A.  1995. Vol.  52. P. 1875 -- 1883.




\bibitem{1998-Gersten}
 Gersten A.
  Maxwell equations as the one-photon quantum equation //
Found.  of Phys. Lett. 1998. Vol.  12. P.   291 -- 298; arXiv:quant-ph/9911049.





\bibitem{1998-Esposito}
 Esposito S.  Covariant Majorana formulation of
electrodynamics //  Found. Phys.   1998. Vol. 28. P.   231 -- 244; arXiv:hep-th/9704144.



\bibitem{1998-Dvoeglazov}
 Dvoeglazov V. Historical note on relativistic theories of
electromagnetism  //   Apeiron.  1998. Vol. 5. P.  69 -- 88.




\bibitem{2001-Ivezic(1)}
 Ivezi\'c T. Lorentz invariant Majorana formulation of the
field equations and Dirac-like Equation for the Free Photon //
EJTP.  2006. Vol.  3. P.   131 -- 142.





\bibitem{2002-Varlamov}
 Varlamov V.  About algebraic foundations of Majorana -- Oppenheimer
quantum electrodynamics and de Broglie -- Jordan neutrino theory of
light //  Ann. Fond.  L. de Broglie.  2003. Vol.  27. P.   273 -- 286.





\bibitem{2002-Khan}
 Khan S.  Maxwell optics: I. An exact matrix
representation of the Maxwell equations in a medium //
arXiv:physics/0205083;   Maxwell optics: II. An exact
formalism //  arXiv:physics/0205084; Maxwell Optics: III.
Applications // arXiv:physics/0205085.





\bibitem{1928-Tetrode}
 Tetrode H.  Allgemein  relativistishe  Quantentheorie  des
Elektrons //  Zeit. Phys.  19828. Bd.  50. S.  336.


\bibitem{1929-Weyl(1)}
 Weyl  H. Gravitation and the electron //  Proc. Nat. Acad. Sci.
Amer. 1929. Vol.  15. P.   323 -- 334;  Gravitation and the
electron // Rice Inst. Pamphlet.  1929. VOl.  16. P.   280 -- 295;
Elektron und Gravitation // Zeit. Phys.  1929. Bd.  56. S. 330 -- 352.



\bibitem{1929-Fock-Ivanenko}
 Fock V.,   Ivanenko D. \"{U}ber eine m\"ogliche geometrische
Deutung der relativistischen Quanten\-theorie //  Zeit. Phys.,
1929. Bd. 54. S. 798 -- 802;
  G\'{e}ometrie   quantique  lin\'{e}aire
et d\'{e}placement parallele //  C. R. Acad. Sci. Paris. 1929. Vol.   188. P.  1470 -- 1472;
 Fock V.  Geometrisierung der Diracschen Theorie des Elektrons //
Zeit. Phys.   1929. Bd.  57. S.  261 -- 277.



\bibitem{Book}
V.M.  Red'kov. Fields in Riemannian space  and the Lorentz group.
Publishing House "Belarusian Science", Minsk, 2009 (in Russian).






\bibitem{1973-Landau}
 L.D. Landau, E.M. Lifshitz.
The theory of the field.   Moskow, 1973 (in Russian).



\bibitem{2009-Red'kov-Tokarevskaya-Ovsiyuk-Spix}
V.M. Red'kov, N.G. Tokarevskaya, E.M.  Ovsiyuk, George J. Spix.
Maxwell equations in Riemannian space-time,  geometry
 effect on material equations
in media.  NPCS, 2009. Vol. 12. No 3.  P. 232--250.




\bibitem{Kratzer-Franz}
A. Kratzer, W. Franz. Transcendent functions. Mockow,  1963 (in Russian).


\end{thebibliography}
\end{document}